\documentclass[prl,aps,showpacs,floatfix,twocolumn]{revtex4}
\usepackage{times}
\usepackage{amsmath,bm,amsfonts}
\usepackage{graphicx}
\usepackage{color}

\begin{document}

\title{Polarization screening and induced carrier density at the interface
  of LaAlO$_3$ overlayer on SrTiO$_3$ (001)}

\author{Yun Li}
\author{Jaejun Yu}
\email[Corresponding author.\ ]{Email: jyu@snu.ac.kr}
\affiliation{Department of Physics and Astronomy, FPRD, Center for
  Strongly Correlated Materials Research, Seoul National University, Seoul
  151-747, Korea}

\date{\today}

\begin{abstract}
  We investigate the role of lattice polarization in determination of
  induced carrier density at the $n$-type interface of LaAlO$_3$ overlayer
  on SrTiO$_3$ (001) by carrying out density-functional-theory
  calculations. When no oxygen vacancy or defect is present, the magnitude
  of polarization screening in the LaAlO$_3$ layers is found to be
  correlated with the carrier charge induced at the interface.  For the
  interfaces with a few LaAlO$_3$ layers, the induced charge carrier is
  compensated by the electrostatic screening and consequently its density
  remains far less than 0.5 electrons per unit cell.
\end{abstract}


\maketitle

The observation of a high mobility electron gas in the $n$-type
(LaO)/(TiO$_{2}$) interface between two band-gap insulators LaAlO$_{3}$
(LAO) and SrTiO$_{3}$ (STO) has generated intense research activities
toward its potential device applications as well as its physical mechanism
\cite{ohtomo04}. The electronic reconstruction at the interface has been
suggested as a way to avoid the polarization catastrophe \cite{nakaga06}
which may arise from the alternating stack of positively charged LaO and
negatively charged AlO$_{2}$ layers on top of the TiO$_{2}$ termination of
STO substrate.

While the reports of ferromagnetism \cite{brinkman07} and
superconductivity \cite{reyren07} in the LAO/STO interface have boosted up
research interest in the mechanism of conductivity and the dimensionality
of the induced charge carrier at the interface, there still remain
controversies on the origin and nature of the interface electron gas
\cite{basletic08,yoshimatsu08}. Several experiments have demonstrated that
oxygen vacancies in the STO layer are responsible for the high carrier
density, which depends more on the film growth and annealing conditions
\cite{kalabukhov07,herranz07,siemons07}. Apart from the
oxygen-vacancy-generated carriers in samples grown in oxygen-poor
conditions, the carrier density of the $n$-type interfaces with perfect
stoichiometry poses another puzzle that a common lower limit in the
carrier density at $\sim 10^{13}$, i.e., 0.03 electrons per unit-cell
(u.c.), is order-of-magnitude less than that of 0.5 electrons per {u.c.}
expected from the electronic reconstruction mechanism \cite{thiel06}.

On the other hand, recent density-functional-theory (DFT) calculations
pointed out the importance of polar distortions as a source of dielectric
screening in the LAO/STO heterostructure
\cite{ishibashi,pentcheva08}. Without the lattice relaxation, even a single
layer of LAO on STO(001) with an ideal structure would become
metallic. Further the insulator-to-metal transition can be driven by an
external electrical field \cite{thiel06,cen08}. The thickness dependence
of carrier density also suggested a possible role of electrostatic
screening in the LAO/STO interface \cite{huijben06}.

Here we show that the detailed balance between the lattice polarization
and the charge transfer is important in determination of the carrier
density at the $n$-type interface of LAO overlayers on STO(001) with
perfect stoichiometry.  Our DFT calculations demonstrate that the lattice
polarization of the LAO overlayer is correlated with the carrier charge
induced at the interface in terms of the overlayer thickness.  When the
LAO overlayers are over a critical thickness, the charge transfer from the
LAO surface to the interface is compensated by the electrostatic screening
due to the polarization distortions across the LAO layers.

We carried out DFT calculations by using the Vienna \textit{ab initio}
Simulation Package (VASP) \cite{kresse96} within a generalized gradient
approximation \cite{wang91} together with the projector augmented wave
pseudopotentials \cite{blochl94,kresse96} and the cut-off energy of 400 eV
for the plane wave basis.  We modeled the LAO/STO interface by a slab
consisting of one-to-ten LAO layers on top of seven STO(001) layers and a
vacuum region of 14 {\AA} along the $c$-axis in a supercell
geometry. Dipole corrections were used to correct the errors of
electrostatic potential, forces, and total energy caused by periodic
boundary condition \cite{makov95}. The in-plane unit cell for calculations
was taken as (1$\times$1) because the rotation distortion of TiO$_{6}$
octahedron is known to be negligible for small carrier doping
\cite{uchida03,pentcheva08}. The in-plane lattice constant of the slab was
constrained at the calculated equilibrium lattice constant $a=3.942$ {\AA}
of the STO substrate. The $c$-axis coordinates of atomic positions were
fully relaxed with forces less than 0.01eV/{\AA} except for the atoms in
the bottom three layers of STO, which were fixed in their bulk structure.

In order to determine accurately the amount and distribution of carrier
density as a function of the layer thickness, we had to devise a proper
k-point sampling for the LAO/STO slab calculations.  For insulating
states, the k-point grid of (12$\times$12$\times$1) was found to be good
enough for the self-consistency iterations. For the case with metallic
charge carriers, however, it was required to introduce an extra dense grid
set near $\Gamma$ and $M$ points because the accurate evaluation of the
tiny pocket sizes of Fermi surfaces is critical for the correct
description of the charge carrier as well as the electrostatic potential
across the interface.

One of the most prominent features in the calculated relaxation geometry
of the LAO/STO slab is the presence of a large polar distortion in the LaO
and AlO$_{2}$ layers but a negligible distortion in the STO side
\cite{pentcheva08}. Because of the huge dipole field produced by the
alternating stack of (LaO)$^{+}$ and (AlO$_{2}$)$^{-}$ charged layers, all
the LaO and AlO$_{2}$ layers become buckled except the surface AlO$_{2}$
layer which exhibits only a small uniform relaxation.  The relative
displacement between La and O ions within the LaO layers varies from 0.1
{\AA} to 0.35 {\AA}, depending on the LAO thickness as well as the
location of LaO layers, while the relative displacements between Al and O
ions within the AlO$_{2}$ layers remain less than 0.15 {\AA}.

The screening by the polar distortion is critical in understanding the
interface electronic structure. We obtained an interesting relaxation
dependence of the LAO/STO interface character in the overlayer structure,
which is similar to the results by Pentcheva and Pickett
\cite{pentcheva08} but apparently quite different from that of
(LAO)$_{m}$/(STO)$_{n}$ superlattices \cite{janicka09}. Regardless of
their thickness, all the \textit{unrelaxed} overlayer structures were
found to have a metallic interface. After the relaxation, however, the
LAO/STO interface remains insulating up to four layers of LAO.  In the
systems with one-to-four LAO layers, the lattice polar distortion is
favored energetically over the charge transfer between the surface and the
interface.

\begin{figure}[htbp]
    \centering
    \includegraphics[width=0.45\textwidth]{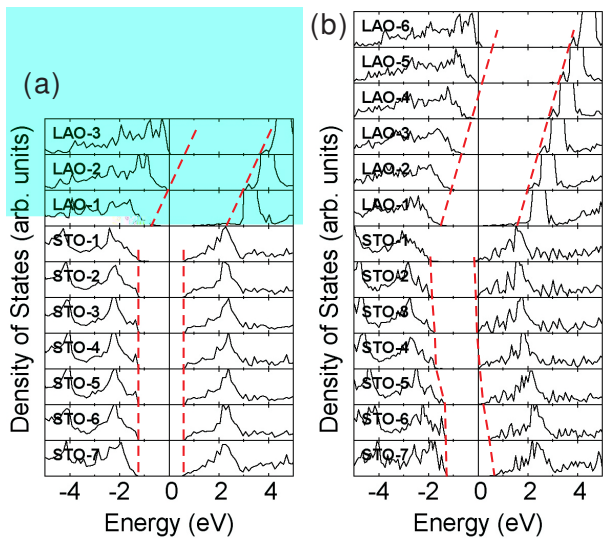}
    \caption{(Color online) Layer projected density-of-states (pDOS) of
      (a) (LAO)$_{3} $/(STO)$_{7}$ and (b) (LAO)$_{6}$/(STO)$_{7}$. Dashed
      lines are drawn by connecting band edges to represent the
      layer-dependent potential profile.} \label{fig:1}
\end{figure}

Figure~\ref{fig:1} shows the layer-projected density-of-states (pDOS) of
two representative systems with fully relaxed LAO layers: (a)
(LAO)$_{3}$/(STO)$_{7}$ and (b) (LAO)$_{6}$/(STO)$_{7}$. The pDOS of the
relaxed (LAO)$_{3}$/(STO)$_{7}$ exhibits an insulating band structure
where the valence band maximum (VBM) of the LAO layers is set at the LAO
surface layer.  The dipole field in the LAO layers shifts the LAO band
edges toward the higher energy relative to those of STO. The potential
gradient is about 0.6 eV per layer for the relaxed structure, whereas it
would be about 1.3 eV for the unrelaxed structure without polar
distortions.  It is noted that the VBM of the LAO surface layers, i.e.,
LAO-3 of Fig.~\ref{fig:1}(a) and LAO-6 of Fig.~\ref{fig:1}(b) deviates
significantly from the extrapolated (dashed) lines connecting the VBM's of
its sub-surface layers. The lowering of the VBM of the LAO surface layer
is attributed to its local environment being close to the bulk
Al$_{2}$O$_{3}$ with a band gap larger than that of LAO. In addition to
the surface effect, there also exists a band offset at the
interface. There is a small but clear offset of about 0.5 eV between the
LAO-1 and STO-1 layers, while the conduction band minimum (CBM) of the
LAO-1 interface layer smears due to the hybridization between LaO and
TiO$_{2}$ layers.

The band edge shift in the LAO side reflects the change of electrostatic
potential across the LAO overlayers. In Fig.~\ref{fig:1}, the
layer-dependent potential profiles are marked by dashed lines.  The
gradient of the edge shift in (a) (LAO)$_{3}$/(STO)$_{7}$ is higher than
that in (b) (LAO)$_{6}$/(STO)$ _{7}$. From our calculations, in the system
with $m=4$ in (LAO)$_{m}$/(STO)$_{7}$, the VBM of the surface LAO layer
barely touches the CBM of the interface STO layer.  When we have more than
four LAO layers, the LAO/STO interface becomes metallic. As shown in
Fig.~\ref{fig:1}(b) of the pDOS of the relaxed (LAO)$_{6}$/(STO)$_{7}$
structure, the charge carrier at the LAO/STO interface is induced as a
result of charge transfer from the LAO surface.  Despite that the
potential gradient is reduced significantly by the polarization screening,
the total potential shift across the LAO layers exceeds the STO band gap
when the overlayer thickness goes over a critical value of 4 LAO
layers. If the VBM of the surface LAO layer lies above the CBM of the
interface STO layer, a charge transfer can occur between the LAO surface
and the LAO/STO interface. From the results, we found that the total shift
of the VBM is always delimited by the STO band gap for all $m\ge 4$. It
implies that the potential shift is somehow compensated by the charge
transfer. The polar distortion occurs in response to a local electric
field, i.e., the potential gradient in the LAO layers, which in turn
depends on the amount of charge transfer. Therefore, in order to determine
the amount of charge transfer, it is necessary to understand the relation
between the lattice polarization and the charge transfer.

In order to quantify the thickness dependence of the lattice polarization,
we calculated the layer-by-layer polarization over the LAO layers as a
functional of the LAO layer thickness. Polarization $P_{\ell}$ of the
$\ell$-th layer can be estimated by
\begin{equation}
  \label{eq:1}
  P_{\ell} = \frac{1}{\Omega_{\ell}} \sum_{i\in \ell} Z^{*}_{\ell i}
  \delta u_{\ell i} ,
\end{equation}
where $Z^{*}_{\ell i}$ and $\delta u_{\ell i}$ are the Born effective
charge and the displacement of the $i$-th ion in the $\ell$-th layer
respectively, and $\Omega_{\ell}$ is the volume of the LAO unit cell in
the $\ell$-th layer. Here we took the values of the Born effective charges
from Ref.~\onlinecite{gemming06}. Since Eq.~(\ref{eq:1}) can be used for
the either La or Al cation-centered cells, we averaged out the
contributions of the oxygens in the planes bounding the elementary unit
cells \cite{nakhmanson05}.

\begin{figure}[htbp]
    \centering
    \includegraphics[width=0.45\textwidth]{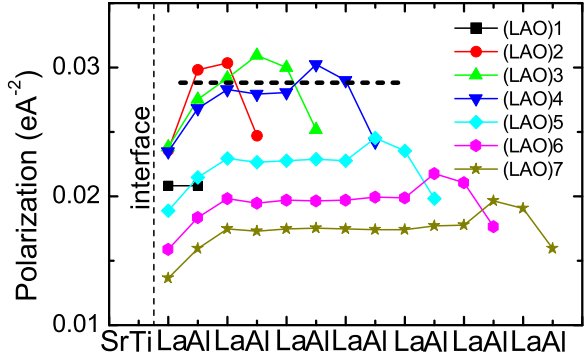}
    \caption{(Color online) Calculated layer-by-layer polarization
    $P_{\ell}$ over the LAO layers for  (LAO)$_{m}$/(STO)$_{7}$
    ($m=1\sim 7$). Labels ``La'' and ``Al'' correspond to the LaO and AlO$_{2}$ layers
    respectively.} \label{fig:2}
\end{figure}

Figure~\ref{fig:2} illustrates the profile of the layer-by-layer
polarization $P_{\ell}$ over the LAO layers for (LAO)$_{m}$/(STO)$_{7}$
($m=1\sim 7$).  In all cases, it is noted that the polarizations of the
interface and surface layers are significantly different from their
average values. Although a reduction of the interface polarization may
arise from both structural relaxation and charge screening, its origin may
require a further investigation but is expected to be different from that
of the ferroelectric insulator superlattices \cite{lee09}.  Aside from the
polarizations at the interface and surface layers, the average values of
$P_{\ell}$ for $n=2$, 3, and 4, denoted by a dashed line in
Fig.~\ref{fig:2}, are roughly constant at about 0.03 e/\AA$^{2}$. From
$m=4$ to 7, The average value of $P_{\ell}$ drops quickly as the layer
thickness increases. This decrease of $P_{\ell}$ is related to the
reduction of the potential gradient in the LAO layers, which is delimited
by the STO band gap and the charge transfer between the surface and the
interface.

\begin{figure}[htbp]
    \centering
    \includegraphics[width=0.45\textwidth]{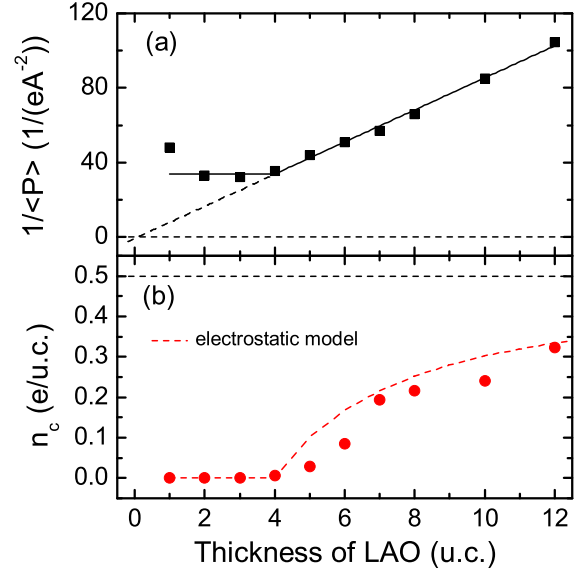}
    \caption{(Color online) The LAO thickness dependence of (a) the average
    inverse polarization $1/\langle P_{\ell}\rangle$ and (b) the charge
    carrier density $n_{c}$.} \label{fig:3}
\end{figure}

For a given thickness $t$ for $m\ge 4$, since the potential shift in the
LAO layers is delimited by the STO band gap, the local electric field
$E_{d}$ can be approximated by $E_{d} = V_{g}^{\mathrm{STO}}/(t-t_{0})$
where $V_{g}^{\mathrm{STO}}$ is a potential change corresponding to the
STO band gap, and $t_{0}$ is an effective parameter corresponding to the dead
layer contribution \cite{stengel06}. As the layer polarization
$P_{\ell}$ is proportional to the local field $E_{d}$, the inverse of the
average polarization $P_{\ell}$ should satisfy the relation:
\begin{equation}
  \label{eq:2}
  \frac{1}{P_{\ell}} = \frac{1}{\chi\epsilon_{0} E_{d}} = \frac{1}{\chi\epsilon_{0}
    V_{g}^{\mathrm{STO}}} (t-t_{0}) ,
\end{equation}
where $\chi$ is an average polarizability of the LAO layers.  The
$t$-dependence of $P_{\ell}$ in Eq.~\eqref{eq:2} is clearly demonstrated
in the plot of ${1}/{\langle P_{\ell}\rangle}$ vs. $t$ as shown in
Fig.~\ref{fig:3}(a).  For $m\ge 4$, $1/\langle P_{\ell}\rangle$ increases
linearly in $t$, while it remains constant below $m=4$ except for the
$m=1$ case. The deviation of the $m=1$ polarization is not surprising
because the single LAO layer consists of both surface and interface
layers.  From the least square fit to the calculated data, we could
determine the effective parameter $t_{0}$ by $t_{0} = 0.053 c_{0}$ where
$c_{0}$ is the unit-cell thickness of the LAO layer.

For the metallic interface with $m\ge 4$, the potential shift, generated
by the dipole field arising from the alternating stack of (LaO)$^{+}$ and
(AlO$_{2}$)$^{-}$ charged layers, is screened by both lattice polarization
and charge transfer. The relation between the interface (or surface)
charge $\sigma_{c}$ and the local electric field $E_{d}$ can be described
in a simple electrostatic picture by the electric displacement field
across the LAO overlayer: $ \sigma_{c} = \epsilon E_{d} =
\epsilon_{0}E_{d} + P $. If there is no free charge carrier at the
interface, an average displacement field should correspond to $\sigma_{c}
= 1/2$ as expected from the polarization catastrophe theory
\cite{nakaga06}. When a charge transfer of $n_{c}$ occurs, however, the
charge screening should give rise to $\sigma_{c} = 1/2 - n_{c}$. By
combining the $t$-dependence of $E_{d}$ and $P_{\ell}$ in
Eq.~\eqref{eq:2}, we obtain an expression for the $t$-dependence of the
induced charge carrier $n_{c}$ at the interface for $m\ge 4$:
\begin{equation}
  \label{eq:3}
  n_{c} = \frac{1}{2} - \frac{A}{t-t_{0}} ,
\end{equation}
where $A = \epsilon_{0}(1+\chi) V_{g}^{\mathrm{STO}}$.  

The calculated charge carriers density $n_{c}$ at the LAO/STO are shown in
Fig.~\ref{fig:3}(b). For $m<4$, the insulating interface has no charge
carrier. At $m=4$, where the VBM of LAO touches the CBM of STO, the
carrier density $n_{c}$ is $\sim 10^{-3}$. For $m>4$, $n_{c}$ increases
monotonically as the thickness of the LAO overlayer grows. For the thin
layers with $m<12$, the carrier densities remain far less than 0.5
e/{u.c.}.  From Eq.~\eqref{eq:3}, it is obvious that $n_{c}=0.5$ e/{u.c.}
can be achieved only when the LAO layers become extremely thick.  By
anchoring the coefficient $A$ to the value of $n_{c}$ at $m=4$, we
obtained $A=1.97$. The fitting result based on the electrostatic model of
Eq.~\eqref{eq:3} is displayed as a dashed line in Fig.~\ref{fig:3}(b).
Despite of some discrepancy for the thin layers close to $m=4$, the
electrostatic model is found to be in a reasonable agreement with the DFT
results.  Considering that the electrostatic model is simplistic and may
not be valid in the scale of a few atomic layers, this agreement is rather
remarkable.

\begin{figure}[htbp]
    \centering
    \includegraphics[width=0.45\textwidth]{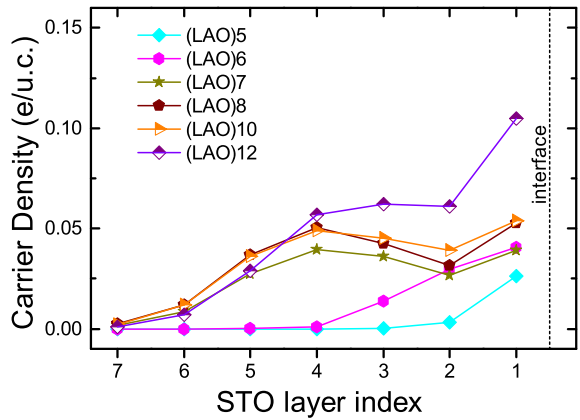}
    \caption{(Color online) Layer-resolved carrier densities in STO for
      the metallic interfaces with the varying thickness of LAO
      overlayers.} \label{fig:4}
\end{figure}

Although our electrostatic model explains the overall behavior of $n_{c}$
without oxygen vacancy, the induced charge carrier still depends on the
detailed electronic structure near the interface. The charge carrier
distribution at the interface is closely related to the band bending,
i.e., the CBM edge shift in the STO substrate. When there is no charge
carrier, i.e., $n_{c}=0$, for the insulating interface with $m<4$, no band
bending occurs in the STO side as shown in Fig.~\ref{fig:1}(a). For the
metallic interface with $m\ge 4$, on the other hand, the degree of the
band bending depends on the amount of the charge carrier. In
Fig.~\ref{fig:4}, we present the layer-resolved carrier distribution as a
function of the LAO layer thickness. The induced charge carriers are found
to be confined within 5 layers close to the LAO/STO interface, while the
distribution for $m=5$ and 6 is even more localized within less than 3
layers.  This result is in good agreement with a recent X-ray
photo-electron spectroscopy (XPS) experiment, which reported that the
carriers far less than 0.5 e/{u.c.}  were found to be confined in one or a
few layers of STO for thin LAO film without oxygen vacancy \cite{sing08}.

In conclusion, we explored a microscopic picture for the screening
mechanism at the interface of the LAO overlayer on STO(001) by carrying
out the DFT calculations, and demonstrated that the lattice polarization
of the LAO layers is correlated with the carrier charge induced at the
interface in terms of the overlayer thickness.  Here we suggest an
electrostatic model for the description of the charge carrier induced at
the interface, where the detailed balance between the lattice polarization
and the charge transfer plays a primary role when no oxygen vacancy or
defect is present in the system.  Although our discussion is restricted to
the perfect lattice without defect, the electrostatic screening mechanism
across the LAO overlayer should be considered as an alternative to the
electronic reconstruction mechanism. We hope that our findings contribute
to resolving the controversies on the origin and nature of the interface
electron gas. In this picture, it is natural to observe a common lower
limit of 0.03 electrons per {u.c.} for the systems with 4-to-5 LAO layers.
In addition, one can consider the induced carrier density controlled by an
external field, which affect the lattice polarization over the LAO layers.

This work was supported by the KOSEF through the ARP
(R17-2008-033-01000-0).  We also acknowledge the support from KISTI under
the Supercomputing Application Support Program.

% \newpage
% \begin{center}
%   Figure Captions
% \end{center}

% \noindent
% Figure 1. (a)
% \medskip

% \noindent
% Figure 2. (Color online)
% \medskip

% \noindent
% Figure 3. (Color online)

\end{document}